\documentclass{article}

\usepackage{arxiv}

\usepackage[utf8]{inputenc} 
\usepackage[T1]{fontenc}    
\usepackage{hyperref}       
\usepackage{url}            
\usepackage{booktabs}       
\usepackage{amsfonts}       
\usepackage{nicefrac}       
\usepackage{microtype}      
\usepackage{lipsum}		
\usepackage{graphicx}
\usepackage[sort&compress,numbers]{natbib}
\usepackage{doi}
\usepackage{chemformula}

\title{Comparative study of Kondo effect in Vanadium dichalcogenides VX$_2$ (X=Se \& Te)}

\author{ \href{https://orcid.org/0000-0002-8478-2948}{\includegraphics[scale=0.06]{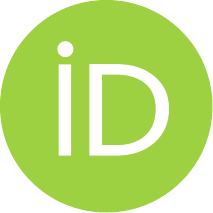}\hspace{1mm}Indrani Kar} \\
	Department of Condensed Matter and Materials Physics\\
	S. N. Bose National Centre for Basic Sciences\\
	Salt Lake, JD Block, Sector III, Bidhannagar, Kolkata-700106, India \\
	\And
	\href{https://orcid.org/0000-0002-0540-0164}{\includegraphics[scale=0.06]{orcid.pdf}\hspace{1mm}Susanta Ghosh} \\
    Department of Condensed Matter and Materials Physics\\
	S. N. Bose National Centre for Basic Sciences\\
	Salt Lake, JD Block, Sector III, Bidhannagar, Kolkata-700106, India \\
	\AND
	\href{https://orcid.org/0000-0001-7762-4004}{\includegraphics[scale=0.06]{orcid.pdf}\hspace{1mm}Shuvankar Gupta} \\
	Condensed Matter Physics Division,\\
	Saha Institute of Nuclear Physics\\
    A CI of Homi Bhabha National Institute\\
	1/AF,Bidhannagar, Kolkata-700064,India\\
	\And
    \href{https://orcid.org/0000-0002-3656-0881}{\includegraphics[scale=0.06]{orcid.pdf}\hspace{1mm}Sudip Chakraborty} \\
	Condensed Matter Physics Division,\\
	Saha Institute of Nuclear Physics\\
    A CI of Homi Bhabha National Institute\\
	1/AF,Bidhannagar, Kolkata-700064,India\\
    \And
    \href{https://orcid.org/0000-0003-1258-0981}{\includegraphics[scale=0.06]{orcid.pdf}\hspace{1mm}S. Thirupathaiah*} \\
	Department of Condensed Matter and Materials Physics\\
	S. N. Bose National Centre for Basic Sciences\\
	Salt Lake, JD Block, Sector III, Bidhannagar, Kolkata-700106, India \\
	\texttt{setti@bose.res.in} \\
}

\renewcommand{\shorttitle}{Kondo effect in VX$_2$ (X=Se \& Te)}

\hypersetup{
pdftitle={A template for the arxiv style},
pdfsubject={q-bio.NC, q-bio.QM},
pdfauthor={David S.~Hippocampus, Elias D.~Striatum},
pdfkeywords={First keyword, Second keyword, More},
}

\begin{document}
\maketitle

\begin{abstract}
We report on the electrical transport, magnetotransport, and magnetic properties studies on the transition metal dichalcogenides VSe$_2$ and VTe$_2$ and draw a comprehensive comparison between them. We observe Kondo effect in both systems induced by the exchange interaction between localized moments and conduction electrons at low temperature, resulting into resistance upturn at 6 K for VSe$_2$ and 17 K for VTe$_2$. From the field dependent resistance measurements we find that the data is fitted best with modified Hamann equation corrected by the quantum Brillouin function for VSe$_2$, while the data is fitted best with modified Hamann equation corrected by the classical Langevin function for
VTe$_2$. Interestingly, we observe a contrasting magnetoresistance (MR) property between these systems across the Kondo temperature. That means, negative MR is found in both systems in the Kondo state. In the normal state MR is positive for VSe$_2$, while it is negligible for VTe$_2$. In addition, both systems show weak ferromagnetism at low temperature due to intercalated V atoms.
\end{abstract}


\section{Introduction}\label{1}
For more than six decades, the TMDCs are widely discussed in the verge of the ground state properties like CDW and superconductivity~\cite{Brixner1962, Brown1966, Nitsche1961, Wilson1969, Wilson1975, Sharma2002, Thorne1996, WAYMAN1996, Kar2020}. Among them, from the group V TMDCs, the ditelluride-based compounds are isostructural to each other having the monoclinic crystal structure with $1T^\prime$ phase below 300 K~\cite{Selte1965, Chen2018b, Clerc2007, Bronsema1984, Liu2016, Soergel2006}, while the diselenide- and the disulfur-based compounds are available in the trigonal crystal structure with $1T$ or $2H$ phase~\cite{Kumakura1996, Terashima2003, Gye2019, Ritschel2015, Leroux2018, Mulazzi2010}.  Despite being isostructural, these compounds show differing electronic and magnetic properties. For instance, \ch{VTe2} shows a normal to commensurate CDW (C-CDW) transition at 474 K~\cite{Bronsema1984, Ohtani1981, Ohtani1984}, while \ch{NbTe2} shows incommensurate CDW (IC-CDW) transition at 550 K and IC-CDW to C-CDW transition at room temperature~\cite{Chen2018b, Battaglia2005}, in addition to a superconducting transition ($T_c$) at $\approx$ 0.5 K~\cite{Nagata1993, VanMaaren1967}. On the other hand, \ch{TaTe2} shows a normal to IC-CDW transition at 170 K~\cite{Chen2017, Chen2018, Liu2015}. Similarly,  \ch{VSe2} shows an IC-CDW transition at around 110 K and an IC-CDW to C-CDW at 70 K~\cite{Thompson1978, Schneemeyer1978, Thompson1979, Pandey2020}. \ch{NbSe2} shows an IC-CDW transition at 33 K and superconductivity at 7.2 K~\cite{Valla2004}. \ch{TaSe2} shows an IC-CDW transition at 600 K and a C-CDW transition at 473 K in the 1T phase~\cite{DiSalvo1975}, while an IC-CDW transition at 122 K and a C-CDW transition at 90 K is found in the 2H phase~\cite{Kumakura1996, Borsa1977}. \ch{VS2} is reported to show a CDW transition below 304 K~\cite{Tsuda1983, Mulazzi2010, Sun2015}. Though \ch{NbS2} is not found to show a clear CDW order, recently it was suggested to be at the verge of a CDW order following the diverse electronic properties in \ch{NbS2}~\cite{Leroux2012a, Leroux2018}. On the other hand, \ch{TaS2} shows IC-CDW phase below 550 K and a nearly commensurate CDW (NC-CDW) phase below 350 K and C-CDW phase was found below 180 K in the 1T phase~\cite{Wang2020}, while a short range CDW transition is found below 75 K in the 2H phase~\cite{Joshi2019}. Magnetotransport studies on \ch{TaTe2} and \ch{NbTe2} suggest a linear dependence of MR on the applied field~\cite{Chen2018b, Chen2017}, while a quadratic MR is found from \ch{VTe2}~\cite{Ding2021}. In contrast to \ch{TaTe2} and \ch{NbTe2}, \ch{VTe2} shows a weak ferromagnetic ordering at low temperature. Also, \ch{VTe2} is found to show a resistivity upturn due to Kondo effect~\cite{Barua2017, Cao2017, Pandey2020, Ding2021}. However, the resistivity upturn found in \ch{VSe2} is still under debate. Some reports interpreted the resistivity upturn to the Kondo effect in the paramagnetic regime~\cite{Barua2017}, while the other reports interpreted it to a weak localization~\cite{Ding2021}.  \ch{TaSe2} in $2H$ phase is reported to show antiferromagnetic ordering~\cite{Lee1970}. \ch{TaS2} in the 1T phase shows ferromagnetic ordering due to localized spins~\cite{Furukawa1985}. As a matter of fact, Kondo effect was not observed in \ch{TaTe2} and \ch{NbTe2} systems.

In this contribution, we report a comprehensive comparison study between \ch{VSe2} and its isovalent compound \ch{VTe2} on their electrical transport, magnetotransport, and magnetic properties. We observe a low temperature resistivity upturn in both systems due to Kondo effect induced by the exchange interaction between localized moments and conduction electrons at low temperature. The electrical transport data of \ch{VSe2} measured under various applied magnetic fields can be explained well with the modified Hamann equation following the Brillouin function~\cite{Liu2019a} using the quantum mechanical behaviour of magnetic moments~\cite{Kittel2005}. From a similar transport study on \ch{VTe2} we find that the data can be fitted well with modified Hamann equation following the Langevin function~\cite{Ding2021} using the classical theory~\cite{Kittel2005}. Foremost these observations confirm the presence of Kondo effect in both systems, specifically for \ch{VSe2} in which the mechanism of resistivity upturn is elusive~\cite{Barua2017, Pandey2020, Ding2021}. Further, we find negative magnetoresistance (MR) from both systems in the Kondo state, while MR behaves differently in the normal state.

\section{Experimental Details}\label{2}
Single crystal of \ch{VTe2} was grown by the chemical vapor transport (CVT) technique using iodine (crystalline, 99.99\%, metals basis, Alfa Aesar) (2 mg/cm\textsuperscript{3}) as a transport agent\cite{SCHAeFER1964}. In the first step, stoichiometric amounts of V (powder, 99.5\%, metals basis, Alfa Aesar) and Te (powder, 99.99\%, metals basis, Alfa Aesar) were mixed thoroughly and sealed in a quartz ampoule under vacuum. The ampoule was then slowly heated to 1000\textsuperscript{$\circ$}C at a rate of 2.5\textsuperscript{$\circ$}C/min and kept there for 8 h before quenched in the normal water. The powder was regrinded, sealed in quartz ampoule under vacuum together with pieces of crystallized iodine (crystalline, 99.99\%, metals basis, Alfa Aesar) (2 mg/cm\textsuperscript{3}). The ampoule was loaded into a three-zone tube furnace where the temperatures were set at 900\textsuperscript{$\circ$}C at the hot-zone and 810\textsuperscript{$\circ$}C was set at the cold-zone. After 5 days of reaction, we obtained shiny single crystals with a typical dimension of 5 mm$\times$5 mm at the cold-zone. Similarly, single crystals of \ch{VSe2} were grown by the above method my mixing the stoichiometric amounts of V (powder, 99.5\%, metals basis, Alfa Aesar) and Se (shot, 99.999\%, metals basis, Alfa Aesar), except that 1000\textsuperscript{$\circ$}C was set at the hot-zone and 950\textsuperscript{$\circ$}C was set at the cold-zone. After 5 days of reaction, we obtained shiny single crystals of typical dimension 15 mm$\times$6 mm at the cold-zone.

Chemical composition of the single crystals were determined by the energy-dispersive X-ray analysis (EDX) equipped with a scanning electron microscope (Quanta 250 FEG) and phase purity was checked by X-ray diffraction (XRD) patterns measured using Cu-k\textsubscript{$\alpha$}-radiation (Rigaku MiniFlex II and Rigaku SmartLab 9KW). Differential scanning calorimetry (DSC) (Q2000 of TA Instruments) was performed to measure the heat flow curve of the sample in both heating and cooling modes.

Electrical transport measurements were done in a physical property measurement system (Quantum Design PPMS-9T) using a standard four-probe method, with the electrical current applied along the plane of sample ($ab$-plane). For magnetotransport measurements the magnetic field was applied at different polar angles with respect to the $ab$-plane. Copper (Cu) leads were connected to the sample by vacuum compatible silver epoxy Epo-Tek H2OE. The sample temperature was varied between 2 K and 330 K during the transport measurements. DC magnetization measurements were performed using the magnetic property measurement system (Quantum Design MPMS-7T) equipped with vibrating sample magnetometer superconducting quantum interference device (VSM-SQUID). Temperature dependence of the magnetization in zero-field-cooled (ZFC) and field-cooled (FC) modes has been carried out under different applied magnetic fields up to 7 T in the temperature range of 5-300 K. Field dependent magnetization [$M(H)$] has also been carried out at different temperatures.

\section{Results and Discussions}\label{3}
\begin{figure}[htb!]
  \centering
  \includegraphics[width=0.5\textwidth]{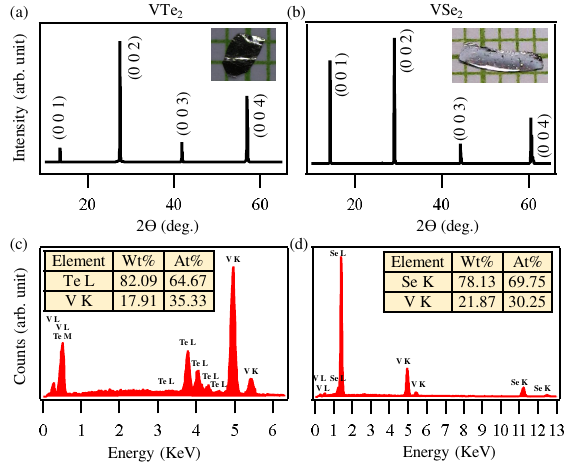}
  \caption{(a) and (b) Powder X-ray diffraction (XRD) pattern of VTe$_2$ and VSe$_2$ single crystals. Insets in (a) and (b) are the optical images of the single crystals. (c) and (d) Energy dispersive X-ray spectroscopy (EDX) data of VTe$_2$ and VSe$_2$ single crystals, respectively.}
  \label{VX1}
\end{figure}

Single crystals of \ch{VTe2} and \ch{VSe2} were structurally analyzed using the powder X-ray diffraction (XRD) at room temperature. High intense diffraction peaks corresponding to the (001) plane of VTe$_2$ are shown in Fig.~\ref{VX1}(a) and (001) plane of VSe$_2$ are shown in Fig.~\ref{VX1}(b). \ch{VTe2} crystallizes into the monoclinic structure with the space group $C2/m$ (12) in the $1T^{\prime}$ phase at room temperature, while \ch{VSe2} crystalizes into the trigonal structure with space group of $P\bar{3}m1$ (164) in the 1$T$ phase~\cite{Bronsema1984, Wiegers1980}. From EDX measurements as shown in Figs.~\ref{VX1}(c) and ~\ref{VX1}(d) we identify that the studied samples have actual compositions of V$_{1.08}$Te$_2$ and V$_{0.86}$Se$_2$, respectively. Thus, we obtained 8\% vanadium excess VTe$_2$ and 14\% vanadium deficient VSe$_2$ single crystals. Herein, for the convenience,  we represent these crystals by their nominal compositions of \ch{VSe2} and \ch{VTe2}.



\begin{figure*} [htb!]
  \centering
  \includegraphics[width=0.95\textwidth]{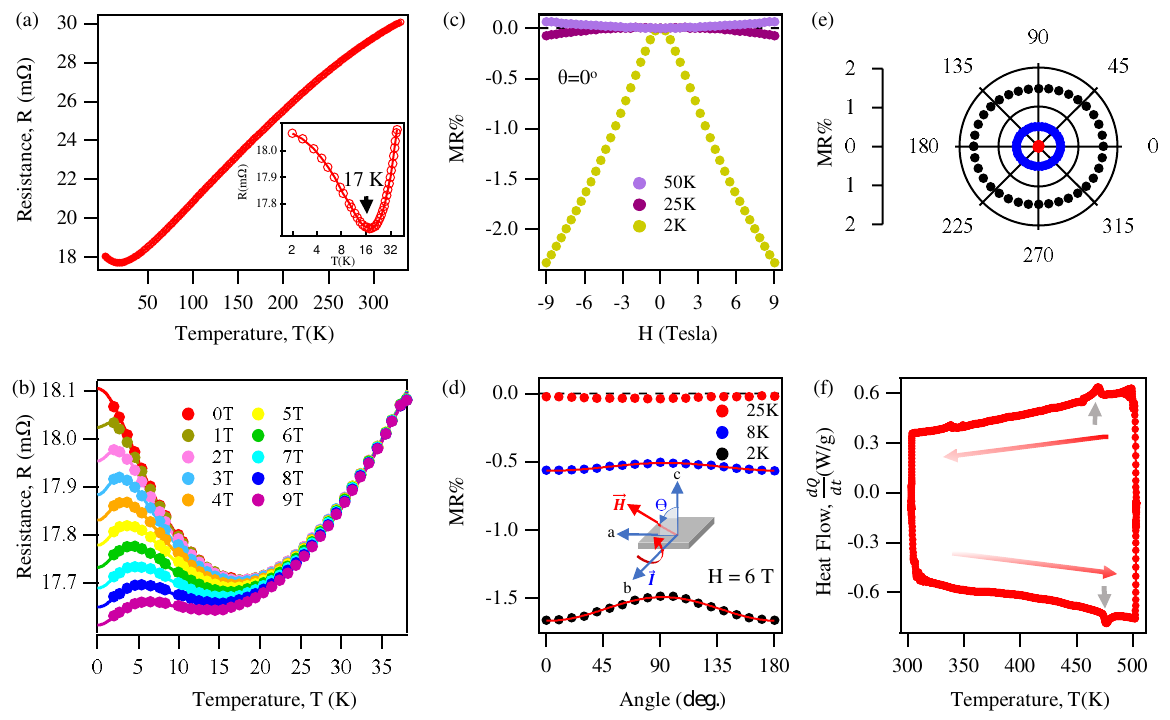}
  \caption{(a) In-plane electrical resistance of VTe$_2$ plotted as a function of temperature. Inset in (a) shows low-temperature resistance upturn at a temperature minima T$_m$=17 K. (b) Temperature dependent electrical resistance plotted for various magnetic fields applied parallel to $c$-axis. Solid lines in (b) are fittings to Eq.~\ref{VXeq1} and Eq.~\ref{VXeq2}. (c) Magnetoresistance, MR (\%),  plotted as a function of field applied parallel to the $c$-axis ($\theta=0^\circ$). (d) MR (\%) plotted as function of applied field angle. Measuring geometry is shown in the inset of (d). (e) Symmetrized MR (\%) of (d) is plotted in the polar graph. (f)   Differential scanning calorimetry (DSC) data is plotted as function of temperature for both cooling and heating cycle. In (f), the charge density wave ordering temperature (T$_{CDW}$) is marked from both heating and cooling cycles.}
  \label{VX2}
\end{figure*}

Fig.~\ref{VX2}(a) shows temperature dependent in-plane electrical resistance of VTe$_2$. The resistance curve shows a metallic nature, in principle, except that an upturn is noticed at low temperature with a resistance minima at T$_m$=17 K as can be seen from the inset of Fig.~\ref{VX2}(a). To explore further on the resistance upturn mechanism, it was measured at various magnetic fields within the temperature range of 2-40 K as plotted in Fig.~\ref{VX2}(b). We observe decrease in resistance upturn with increasing magnetic field of up to 9 T. Further, the field dependent resistance curves are perfectly fitted with the Eq.~\ref{VXeq1} and Eq.~\ref{VXeq2} without and with applied magnetic field, respectively. In Eq.~\ref{VXeq1}, the first term ($R_{0}$) is residual resistance, the second term ($qT^2$) represents the electron-electron interaction, and the third term is the Hamann expression~\cite{Hamann1967} due to $s-d$ exchange interaction~\cite{Yosida1957, Ding2021}. Hamann expression is an empirical equation used to calculate the exchange interaction between localized magnetic moments and conduction electrons. In the third term, $R_{KO}$ is the temperature-independent Kondo resistance, $T_K$ is the Kondo temperature, and S is the total spin of the magnetic impurities. $T_{eff}=\sqrt{T^2+{T_W}^2}$, where ${k_B}T_W$ is the effective Ruderman-Kittel-Kasuya-Yosida (RKKY) interaction strength. From the fitting, we obtained a Kondo temperature $T_K$=12 K and S=0.5. These values are in good agreement with previous report on \ch{VTe2} single crystal~\cite{Ding2021}. In Eq.~\ref{VXeq2}, the Hamann term is multiplied by $[1-L^2(\frac{{\mu}H}{{k_B}{T_eff}})]$ where $L(x)$ is Langevin function $L(x)=\coth(x)-1/x$ (see Table~\ref{VXT1} for the fitting parameters). Primarily, Hamann equation involves Brillouin function~\cite{Liu2019a} to describe the quantum mechanical behavior of magnetic moments and provides a more accurate description of their statistical distribution. In quantum theory, the magnetic moments are quantized and the orientations of magnetic moment with respect to applied magnetic field are specified by some possible components of magnetic moment along field direction~\cite{Kittel2005}. On the other hand, the Langevin function~\cite{Ding2021} is a simplified model derived from classical statistical mechanics. In Langevin's classical theory, it is assumed that the mutual interaction between the magnetic dipoles is negligible and magnetic moments can possess any orientation with respect to applied magnetic field~\cite{Kittel2005}.

\begin{table}[htb!]
  \centering
\begin{equation}\label{VXeq1}
  R(T)=R_{0}+qT^2+R_{KO}[1-\frac{\ln{(\frac{T_{eff}}{T_K})}}{\sqrt{\ln^{2}{(\frac{T_{eff}}{T_K})}+S(S+1)\pi^2}}]
\end{equation}
\end{table}

\begin{multline}\label{VXeq2}
  R(T)=R_{0}+qT^2+R_{KO}[1-\frac{\ln{(\frac{T_{eff}}{T_K})}}{\sqrt{\ln^{2}{(\frac{T_{eff}}{T_K})}+S(S+1)\pi^2}}][1-L^2(\frac{{\mu}H}{{k_B}{T_eff}})]
\end{multline}

\begin{table}[htb!]
  \centering
    \begin{tabular}{|p{.05\textwidth}| p{.1\textwidth}| p{.1\textwidth}| p{.1\textwidth}| p{.1\textwidth}| p{.1\textwidth}|}
    \hline

      H (T) & $R_0$ ($m\Omega$) & $q$ ($\mu\Omega/K^{2}$) & $R_{KO}$ ($m\Omega$) & $T_W$ (K) & $\mu$ ($\mu_B$) \\
  \hline

   0 & 16.6140(4) & 0.5719(4) & 1.0782(4) & 3.900(3) &  \\
  \hline

  1 & 16.6530(4) & 0.5645(4) & 1.0316(4) & 2.423(3) & 4.042(3) \\
  \hline

  2 & 16.6563(4) & 0.5641(4) & 1.0295(4) & 2.665(3) & 2.569(3) \\
  \hline

 3 & 16.6786(4) & 0.5596(4) & 1.0028(4) & 2.823(3) & 2.045(3) \\
  \hline

   4 & 16.7015(4) & 0.5561(4) & 0.9740(4) & 3.147(3) & 1.789(3) \\
  \hline

  5 & 16.7343(4) & 0.5508(4) & 0.9331(4) & 3.396(3) &  1.631(3) \\
  \hline

  6 & 16.7778(4) & 0.5435(4) & 0.8780(4) & 3.632(3) & 1.510(3) \\
  \hline
    7 & 16.8104(4) & 0.5398(4) & 0.8334(4) & 3.916(3) & 1.445(3) \\
  \hline

   8 & 16.8592(4) & 0.5320(4) & 0.7692(4) & 4.135(3) & 1.379(3) \\
  \hline

   9 & 16.9061(4) & 0.5263(4) & 0.7054(4) & 4.332(3) & 1.329(3) \\
  \hline

  \end{tabular}
 \caption{Kondo fitting parameters of VTe$_2$}\label{VXT1}
\end{table}

Fig.~\ref{VX2}(c) depicts magnetoresistance (MR\%), $MR\%=\frac{R(H)-R(0)}{R(0)} \times 100$,  plotted as a function of applied field measured at below (2 K) and above (25 K \& 50 K) the Kondo temperatures. We notice that at 2 K, below the Kondo temperature, VTe$_2$ shows a negative MR and above Kondo temperature, the MR is negligible. This is because the magnetic field reduces fluctuations of localized impurity magnetic moments and spin dependent exchange scattering~\cite{Sekitani2003}. Hence, below a critical magnetic field (B$_c$), the MR is negative and above B$_c$, the Kondo effect vanishes. In case of VTe$_2$ the value of B$_c$ is very high~\cite{Ding2021} which is beyond our experimental scope. Fig.~\ref{VX2}(d) shows MR (\%) plotted as a function of field angle with respect to the sample surface normal ($c$-axis) at an applied field of 6 T. Measuring geometry is shown in the inset of Fig.~\ref{VX2}(d). Red solid-lines in Fig.~\ref{VX2}(d) are fittings to the Eq.~\ref{VXeq3}. In Eq.~\ref{VXeq3}, the first tern is a constant and $\alpha$ is an amplitude.  Fig.~\ref{VX2} (e) shows symmetrized angular dependent MR plotted in the polar graph for the temperatures 2 K, 8 K,  and 50 K.  As can be seen from Fig.~\ref{VX2} (e), MR shows in-plane small anisotropy in the Kondo state (2 K) but becomes completely isotropic in the normal state (8 and 50 K). Ideally Kondo systems show isotropic MR but a small anisotropy could be there due to the presence of a crystal anisotropy~\cite{Sekitani2003}. DSC measurements are performed in both heating and cooling modes with a ramp rate of dT/dt=10 K/min as depicted in Fig.~\ref{VX2}(f). During cooling and heating cycles of DSC measurements we noticed a hump and a dip, respectively, at a sample temperature of 470$\pm$5 K that is originated from the charge density wave (CDW)~\cite{Ohtani1981}.

\begin{table}[htb!]
\begin{equation}\label{VXeq3}
\centering
MR(\theta)=MR_0+{\alpha}cos{(2\theta)}
\end{equation}
\end{table}

\begin{figure*}[!htb!]
  \centering
  \includegraphics[width=0.9\textwidth]{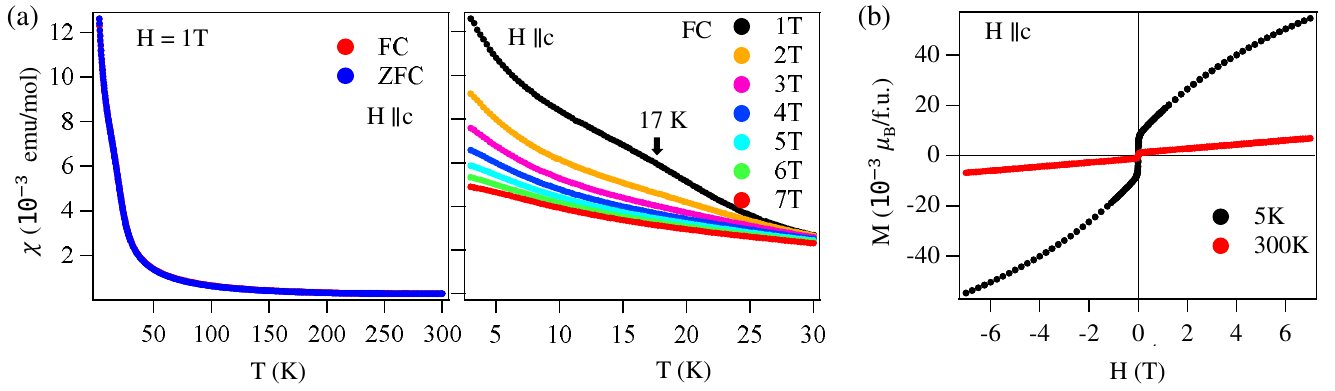}
  \caption{(a) Magnetic susceptibility of VTe$_2$ plotted as a function of temperature with a field of 1 T applied parallel to $c$-axis measured in the FC and ZFC modes.  (b) Same as (a) but measured for different magnetic fields in the FC mode. (c) Magnetization isotherms $M(H)$ measured at 5 K and 300 K.}
  \label{VX3}
\end{figure*}

Fig.~\ref{VX3} shows magnetic measurements on VTe$_2$ with field applied parallel to the $c$-axis. Fig.~\ref{VX3}(a) shows temperature dependent magnetic susceptibility measured with a field of 1 T within the temperature range of 3-300 K in the field-cooled (FC) and zero-field-cooled (ZFC) modes. Here, it can be seen that ZFC and FC are identical. Further, we measured the temperature dependent magnetic susceptibility by varying the field in the FC mode as shown in Fig.~\ref{VX3}(b). A broad hump-like structure with a maximum at around 17 K has been noticed in the susceptibility data when measured with a magnetic field of 1 T, which is then eventually disappears above 3 T as we increase the field due to the crossover from high temperature localized moments of Kondo impurities to fully compensated moments at low temperature~\cite{Park2011, Ding2021}. However, at higher magnetic fields this crossover gets weaker as a result the resistance upturn decreases. Fig.~\ref{VX3}(c) shows magnetization isotherms, M(H), measured at 5 and 300 K. From the M(H) data we notice a soft ferromagnetic-like ordering at the room temperature which is in agreement with previous reports on \ch{VTe2}~\cite{Pan2014, Fuh2016, Liu2019a}.

\begin{figure*}[htb!]
  \centering
  \includegraphics[width=0.95\textwidth]{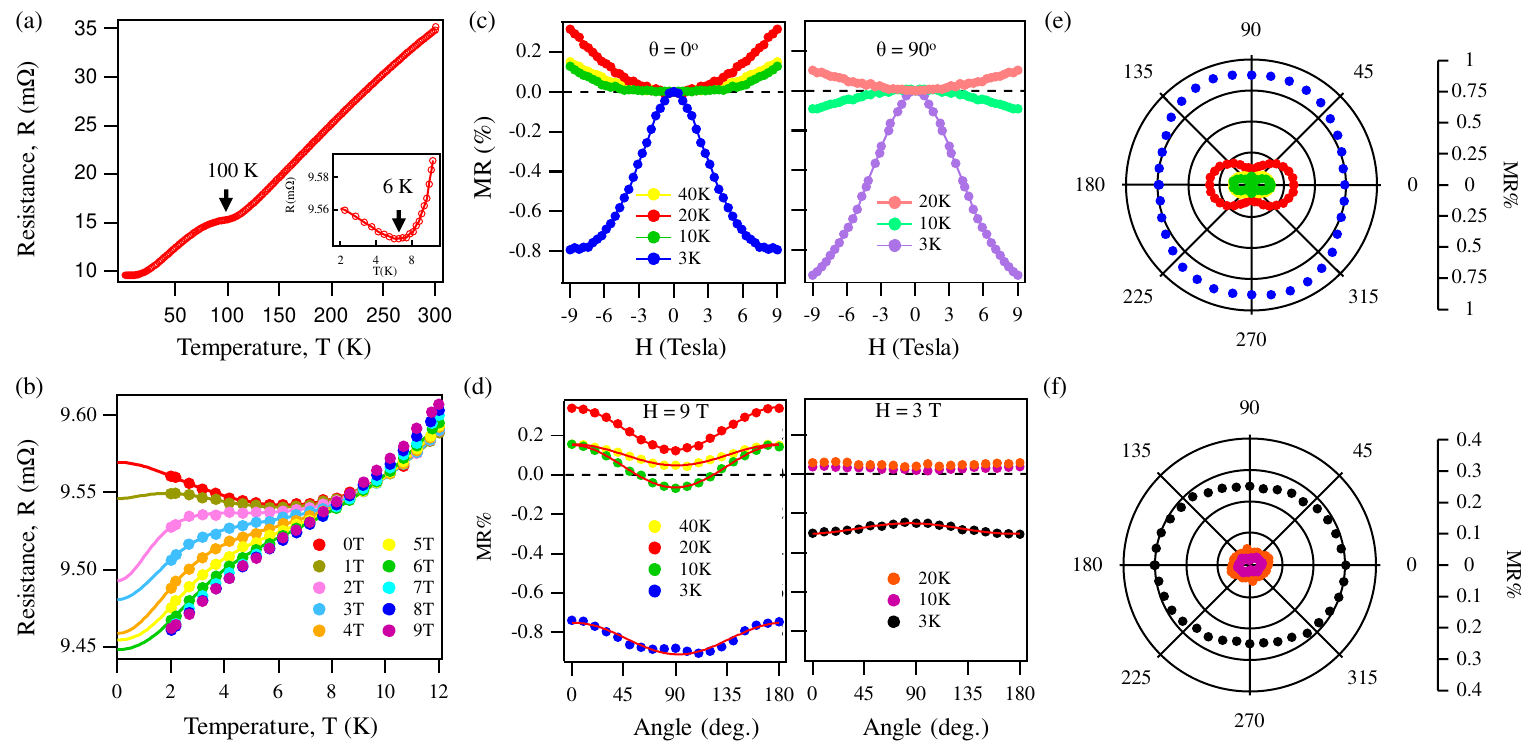}
  \caption{(a) In-plane electrical resistance of VSe$_2$ plotted as a function of temperature. A hump at CDW transition temperature of 100 K is noticed on the resistance curve. Inset in (a) shows the resistance upturn at a temperature minima T$_m$=6 K. (b) Temperature dependent electrical resistance plotted for various magnetic fields applied parallel to the $c$-axis. Solid lines in (b) are fittings to Eq.~\ref{VXeq4} and Eq.~\ref{VXeq5}. (c) Magnetoresistance, MR (\%),  plotted as a function of field applied parallel to the $c$-axis ($\theta=0^\circ$) is in the left panel and for the field applied perpendicular to the $c$-axis ($\theta=90^\circ$) is shown in the right panel. (d) MR (\%) \textbf{is} plotted as function of the field angle with applied field of 9 T (left panel) and 3 T (right panel). (e) Symmetrized MR (\%) of (d) is plotted in the polar graph with an applied field of 9 T. (f) Symmetrized MR (\%) of (d) is plotted in the polar graph with an applied field of 3 T.}
  \label{VX5}
\end{figure*}

Next, Fig.~\ref{VX5}(a) shows temperature dependent zero-field in-plane electrical resistance of VSe$_2$. Overall, the resistance curve suggests a metallic behaviour except that a hump at around 100 K and a resistance upturn at low temperature is observed. The hump at 100 K is due to the incommensurate charge density order (IC-CDW) as noticed in the previous reports from this system~\cite{Cao2017, Barua2017, Pandey2020, Sayers2020, Fumega2019, Toriumi1981, Bruggen1976, Bayard1976, Thompson1978}. Inset in Fig.~\ref{VX5}(a) shows a resistance minima (T\textsubscript{m}) at 6 K. To understand the resistance upturn mechanism, we performed field dependent resistance measurements within the temperature range of 2 K to 12 K by varying the magnetic fields up to 9 T as plotted in Fig.~\ref{VX5}(b). As can be seen from Fig.~\ref{VX5}(b),  resistance upturn decreases with increasing field and disappears at 7 T, suggesting Kondo effect as the possible origin of the resistance upturn. Solid-lines in Fig.~\ref{VX5}(b) are the fittings to Eq.~\ref{VXeq4} in absence of applied magnetic field and Eq.~\ref{VXeq5} in presence of applied magnetic field with a modified Hamann term~\cite{Liu2019a, Barua2017, Kondo1999, Cao2017}. Eq.~\ref{VXeq4} is almost similar to Eq.~\ref{VXeq1}, except that the second term (qT$^3$) in Eq.~\ref{VXeq4} represents the interband ($s-d$) $e-ph$ scattering while it is $e-e$ scattering (qT$^2$) contribution in Eq.~\ref{VXeq1}. From the fitting,  we estimated a Kondo temperature (T$_K$) of about 6 K and the spin component S=0.5, which are in good agreement with previous report on \ch{VSe2}~\cite{Barua2017}. In Eq.~\ref{VXeq5}, the Hamann expression is modified using the Brillouin function, $B(x)=({\frac{2S+1}{2S}}){\coth{{\frac{2S+1}{2S}}}x}-{\frac{1}{2S}}{\coth{{\frac{1}{2S}}}x}$ in the quantum limit. See Table~\ref{VXT2} for the fitting parameters.

\begin{table}[htb!]
 \begin{equation}\label{VXeq4}
  R(T)=R_{0}+qT^3+R_{KO}[1-\frac{\ln{(\frac{T_{eff}}{T_K})}}{\sqrt{\ln^{2}{(\frac{T_{eff}}{T_K})}+S(S+1)\pi^2}}]
\end{equation}
\end{table}

\begin{multline}\label{VXeq5}
  R(T)=R_{0}+qT^3+R_{KO}[1-\frac{\ln{(\frac{T_{eff}}{T_K})}}{\sqrt{\ln^{2}{(\frac{T_{eff}}{T_K})}+S(S+1)\pi^2}}][1-B^2(\frac{{g\mu_B}SH}{{k_B}T_{eff}})]
\end{multline}

Left panel in Fig.~\ref{VX5}(c) depicts the field dependent MR(\%),  measured at various temperatures for the field applied parallel to $c$-axis ($\theta=0^\circ$). The right panel in Fig.~\ref{VX5}(c) depicts MR(\%) measured perpendicular to $c$-axis ($\theta=90^\circ$). We can see from Fig.~\ref{VX5}(c) that for both field orientations the MR is negative (positive) below (above) the Kondo temperature (6 K). Fig.~\ref{VX5}(d) depicts the angle dependent MR(\%) measured at various temperatures with a magnetic field of 9 T (left panel) and 3 T (right panel). From  Fig.~\ref{VX5}(d) we can see that the MR(\%) has a maximum for H$\parallel$c and minimum for H$\perp$c under the magnetic field of 9 T. On the other hand it is reversed  when the applied field is 3 T, which means that the MR(\%) is minimum for H$\parallel$c and maximum for H$\perp$c. Nevertheless, MR (\%) sinusoidally depends on the field angle in going from  H$\parallel$c to  H$\perp$c. The Red solid lines are the fits to Eq.~\ref{VXeq3}. Figs.~\ref{VX5} (e) and ~\ref{VX5} (f) represents the polar maps of the symmetrized data shown left and right panels of Fig.~\ref{VX5} (d), respectively. Figs.~\ref{VX5} (e) and ~\ref{VX5} (f) clearly show the out-of-plane anisotropic MR(\%) in \ch{VSe2}. Interestingly, the anisotropy is rotated by 90$^\circ$ between 3 T and 9 T. Anisotropic MR at 9 T is in good agreement with previous study~\cite{Toriumi1981}. Also, note here that the anisotropic MR is significant above the Kondo state, i.e. at higher magnetic fields (9 T) irrespective of the temperature. In the Kondo state (3 T), a small anisotropic behaviour is observed which is negligible.

\begin{table}[htb!]
  \centering
  \begin{tabular}{|p{.05\textwidth}| p{.1\textwidth}| p{.1\textwidth}| p{.1\textwidth}| p{.1\textwidth}|}
    \hline

  H (T) & $R_0$ ($m\Omega$) & $q$ ($\mu\Omega/K^{3}$) & $R_{KO}$ ($m\Omega$) & $T_W$ (K) \\
  \hline

   0 & 9.4223(4) &  0.0483(4) & 0.1121(4) & 2.455(3)  \\
  \hline

  1 & 9.3600(4) & 0.0485(4) & 0.1751(4) & 2.023(3) \\
  \hline

  2 & 9.4211(4) & 0.0480(4) & 0.1124(4) & 1.452(3) \\
  \hline

  3 & 9.4017(4) & 0.0497(4) & 0.1367(4) & 2.160(3) \\
  \hline

  4 & 9.4301(4) & 0.0497(4) & 0.1061(4) & 1.832(3) \\
  \hline

   5 & 9.4508(4) & 0.0480(4) & 0.0843(4) & 1.385(3) \\
  \hline

 6 & 9.4440(4) & 0.0506(4) & 0.0965(4) & 1.694(3) \\
  \hline

  \end{tabular}
 \caption{Kondo fitting parameters of VSe$_2$}\label{VXT2}
 \end{table}

\begin{figure*}[htb!]
  \centering
  \includegraphics[width=0.95\textwidth]{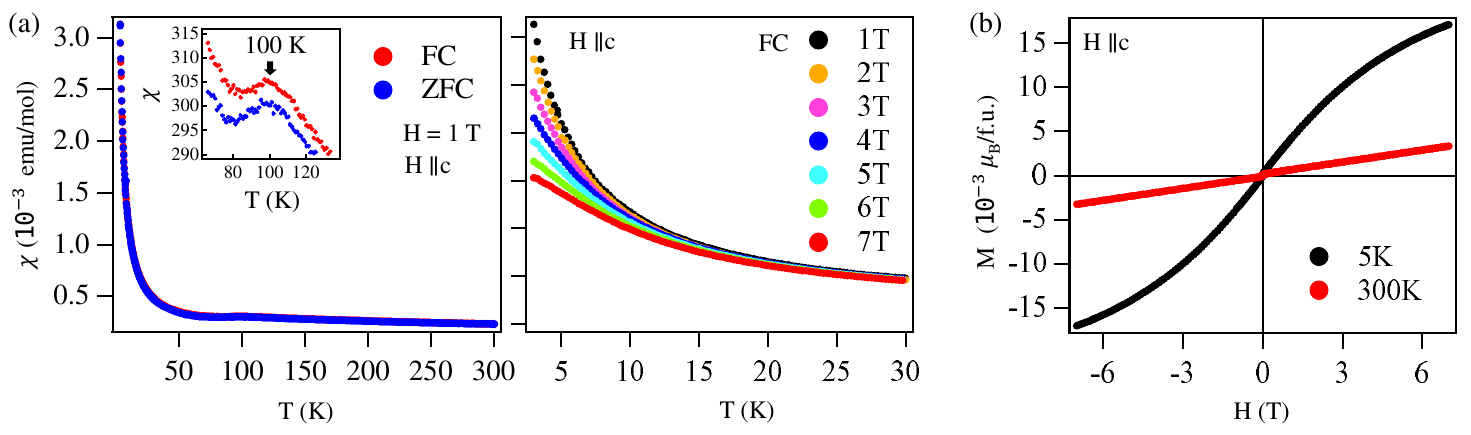}
  \caption{(a) Magnetic susceptibility of VSe$_2$ plotted as a function of temperature measured with a field of 1 T applied parallel to the $c$-axis in the FC and ZFC modes. (b) Same as (a) but measured for different magnetic fields in the FC mode. A hump in the magnetic susceptibility is noticed both in FC and ZFC modes at a CDW transition temperature of 100 K.  (c) Magnetization isotherms $M(H)$ measured at 5 K and 300 K.}
  \label{VX6}
\end{figure*}

Figure~\ref{VX6} shows magnetic properties studies on the VSe$_2$ single crystals. Fig. ~\ref{VX6}(a) shows temperature dependent magnetic susceptibility measured with a magnetic field of 1 T applied parallel to the $c$-axis within the temperature range of 3-300 K in the ZFC and FC modes. Here, it can be seen that ZFC and FC are identical. To observe the effect of Kondo screening on the magnetic properties, we measured the susceptibility by varying the magnetic field  of up to 7 T within the temperature range of 3-30 K in FC mode as shown in Fig. ~\ref{VX6}(b). Unlike in \ch{VTe2}, we do not find a significant change in the susceptibility of \ch{VSe2} as a function of applied field, probably due to the low Kondo temperature (6 K) found in \ch{VSe2}.  Fig.~\ref{VX6}(b) shows the magnetization isotherm $M(H)$ measured at 5 and 300 K. We clearly notice a sigmoid like $M(H)$ curve at 5 K due to soft ferromagnetic ordering. This observation is different from previous reports where \ch{VSe2} is found in bulk to be paramagnetic at all temperatures~\cite{Bonilla2018, Bruggen1976, Bayard1976}. However, a strong ferromagnetic ordering is reported in monolayer VSe$_2$ grown on HOPG or \ch{MoS2} substrate at the room temperature~\cite{Bonilla2018}. Another recent report also demonstrated a small ferromagnetic signal in stoichiometric bulk \ch{VSe2}, which they suggest to arise from vanadium impurities intercalated between the van der Waals layers.

Finally, on comparing the electrical resistivity, magnetotransport, and magnetic properties between \ch{VTe2} and \ch{VSe2},  we find that both systems show low temperature resistivity upturn due to the Kondo effect with a Kondo temperature of 12 K for the former and 6 K for the later, suggesting stronger magnetic correlations in \ch{VTe2} compared to \ch{VSe2}. Both systems show negative MR in the Kondo state. \ch{VSe2} shows negative to positive MR switching across the Kondo temperature of 6 K, while \ch{VTe2} shows negligible MR above the Kondo temperature of 12 K. Moreover, above the Kondo state, a strong anisotropic MR is observed in the case of \ch{VSe2}, while it is almost negligible in \ch{VTe2}. In case of \ch{VTe2} a weak anisotropic MR in the Kondo state is observed, whereas above the Kondo state the MR is negligible. Interestingly, though both systems show a low-temperature resistivity upturn due to the Kondo effect, under the magnetic field it behaves differently for different compounds. That means, \ch{VTe2} follows the Hamann law modified with the classical Langevin function with dominant contribution from the $e-e$ scattering, while \ch{VSe2} follows the Hamann law modified with the quantum Brillouin function (see Eq.~\ref{VXeq5}) with the dominant contribution from the interband ($s-d$) $e-ph$ scattering. Different functions applicable to the same Kondo effect observed in  different compounds can be understood from their differing Kondo temperatures. That means, the Kondo temperature of \ch{VTe2} (12 K) is almost two times higher than the Kondo temperature of \ch{VSe2} (6 K). On the other hand, a previous study on the single-crystalline nanoplates of \ch{VTe2} reported a Kondo temperature of 6 K and the resistivity data was best fitted following the Brillouin function~\cite{Liu2019a}. These observations suggest that the Kondo temperature of these systems is thickness sensitive~\cite{Yang2014}.

\section{Conclusions}\label{4}
In conclusion, we have drawn a comprehensive comparison between the transition-metal dichalcogenides \ch{VSe2} and \ch{VTe2} on their electrical transport, magnetotransport, and magnetic properties. We observe the Kondo effect in both systems induced by the exchange interaction between localized moments and conduction electrons at low temperature, resulting into resistance upturn at Kondo temperature of 6 K for \ch{VSe2} and 12 K for \ch{VTe2}. From the field dependent resistance measurements we find that the data is fitted best with modified Hamann equation corrected by the quantum Brillouin function for \ch{VSe2}, while the data is best fitted with modified Hamann equation corrected by the classical Langevin function for \ch{VSe2}. Interestingly, we observe a contrasting magnetoresistance (MR) properties between these systems across the Kondo temperature. In both systems examined, there is a manifestation of weak ferromagnetism at low temperatures. This phenomenon can be attributed to the presence of intercalated V atoms in both systems.


\section{Acknowledgements}\label{5}
S. T. greatly acknowledges the financial support given by SERB-DST through the Grant no. SRG/2020/00393. This research is made use of the Technical Research Centre (TRC) Instrument Facilities of S. N. Bose National Centre for Basic Sciences, established under the TRC project of Department of Science and Technology, Govt. of India. The authors thank Prof. Priya Mahadevan for fruitful discussions.

\bibliographystyle{unsrt}
\bibliography{Thesis6}  






\end{document}